\documentclass[aps, onecolumn,prb]{revtex4}
\usepackage{graphicx}

\begin{document}

\title{Time domain analysis of dynamical switching in a Josephson junction}
\author{Joachim Sj\"ostrand${}^1$, Jochen Walter${}^2$, David Haviland${}^2$, Hans Hansson${}^1$ and Anders Karlhede${}^1$}
\affiliation{${}^1$Department of Physics, Stockholm University \\
${}^2$Nanostructure Physics, Royal Institute of Technology \\
AlbaNova University Center, SE-10691 Stockholm, Sweden}

\date{\today}


\begin{abstract}
\noindent
We have studied the switching behaviour of a small capacitance Josephson junction both in experiment, and by numerical simulation of a model circuit. The switching is a complex process involving the transition between two dynamical states of the non-linear circuit, arising from a frequency dependent damping of the Josephson junction. We show how a specific type of bias pulse-and-hold, can result in a fast detection of switching, even when the measurement bandwidth of the junction voltage is severely limited, and/or the level of the switching current is rather low.  
\end{abstract}

\maketitle

\vskip0.1pc

\section{Introduction}
\noindent
Josephson junction physics is presently experiencing a renaissance, where many new and exciting experiments on small capacitance Josephson junction circuits are demonstrating quantum dynamics with measurements in the time domain.\cite{Nakamura,Vion,Chirescu,Martinis,Duty} In this field, an engineering approach to the design of quantum circuits is presently being applied to the long term goal of realizing a solid-state, scalable technology for implementation of a quantum bit processor. The intrest in quantum computation is fueled by a theoretical dream of massively parallel computation with quantum two-level systems, or qubits (for a nice introduction, see the book by Nielsen and Chuang \cite{NC}). A central question for these Josephson junction experiments relates to the optimal design for qubit readout, or detection of the quantum state of the circuit. Quantum state readout has been succesfully performed by switching current measurements.

In this paper we use numerical simulation to analyze a particular type of readout--the switching of a Josephson junction subject to frequency dependent damping. The system we model is motivated by experiments, and comparisons are made with measured data. The switching process is a complex transition between two dynamical states of the circuit, and our numerical simulations allow us to study the speed and resolution of the switching process, so that we can investigate the use of this switching process as a detector for quantum state readout of a Cooper Pair Transistor (CPT).

The classical description of the Josephson junction reduces to an analysis of the non-linear dynamics of the Josephson phase variable, which is conveniently discussed in terms of a ficticious "phase particle" moving in a washboard or $\cos(\phi)$ Josephson potential.\cite{Tink} The time evolution of the phase variable depends entirely upon the particular circuit in which the junction is embedded. The simple and most tractable model is the so-called resistively and capacitively shunted junction model (RCSJ) of Stewart\cite{Ste68} and McCumber\cite{McC68}, where an ideal Josephson tunneling element is connected in parallel with an ideal capacitor (the tunnel junction capacitance), an ideal frequency independent impedance (a resistor), and driven by an external current source. More difficult to describe, and more relevant to experiments with small capacitance Josephson junctions, is the phase dynamics when the parallel impedance becomes frequency dependent, and when fluctuations due to the finite temperature of the dissipative elements are included. With small capacitance Josephson junctions we are often in a situation where the frequency dependent shunting impedance gives overdamped phase dynamics at high frequency, and underdamped phase dynamics at low frequency. The generic features of such a model have been studied extensively with computer simulations by e.g. Kautz and Martinis.\cite{KM90} In this model the system can be in two distinct dynamical states for some given bias currents. In the "phase diffusion" state, a small finite voltage appears over the junction when the noise term causes a diffusive motion of the phase particle in a tilted washboard potential. In the "free running" state, a much larger voltage is sustained across the junction. A switching of the junction, or transition between these two dynamical states, occurs when a critical velocity of the phase particle is reached, so that it can overcome a "dissipation barrier" resulting from the frequency dependent damping \cite{Tink,KM90,Vion2,Joyez}. In the present work we study the probability for this dynamical switching, as a function of time and amplitude, for a particular type of bias pulse.

\begin{figure}[tbp]
\begin{center}
\includegraphics[angle=0,width=0.43\textwidth]{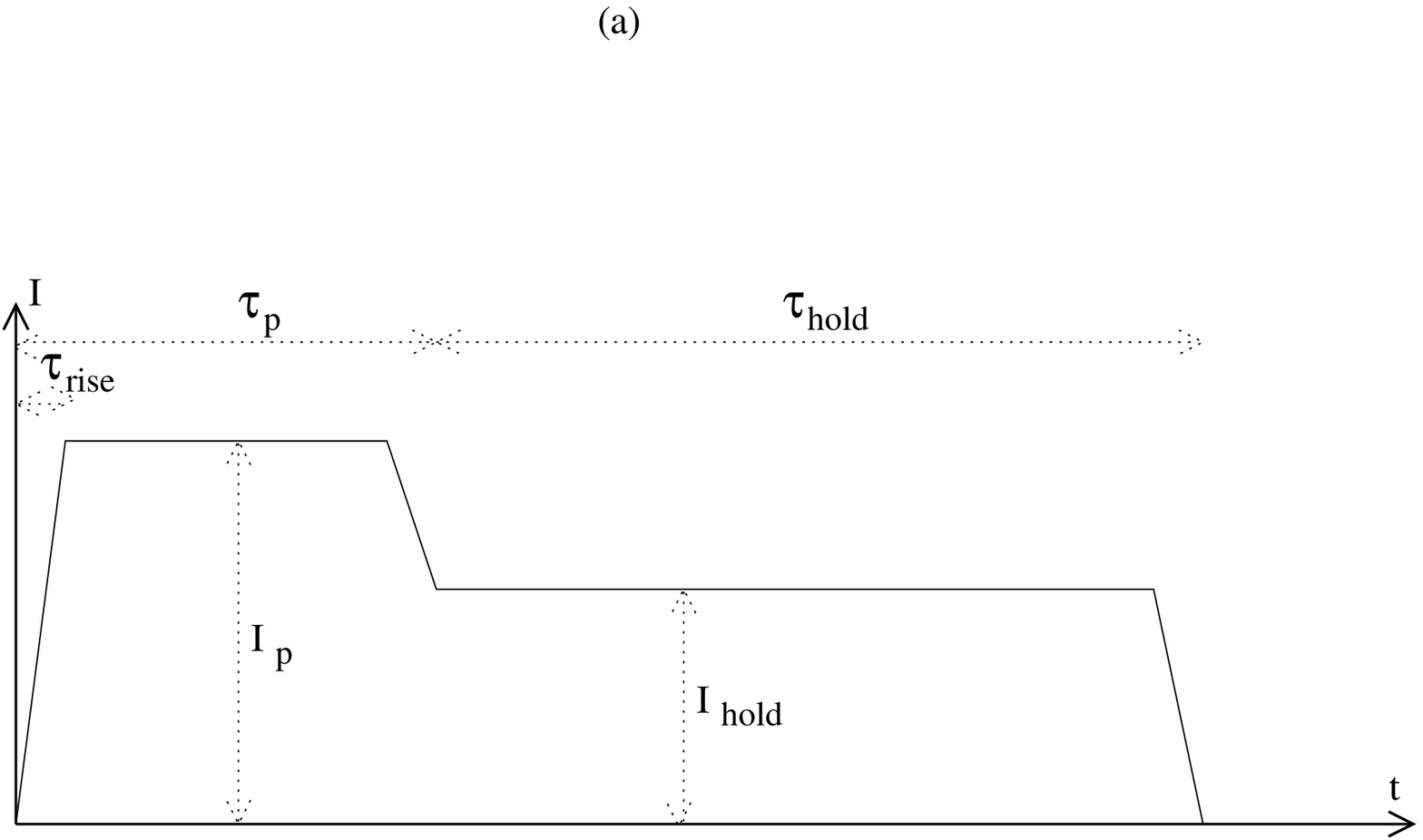}
\includegraphics[angle=0,width=0.40\textwidth]{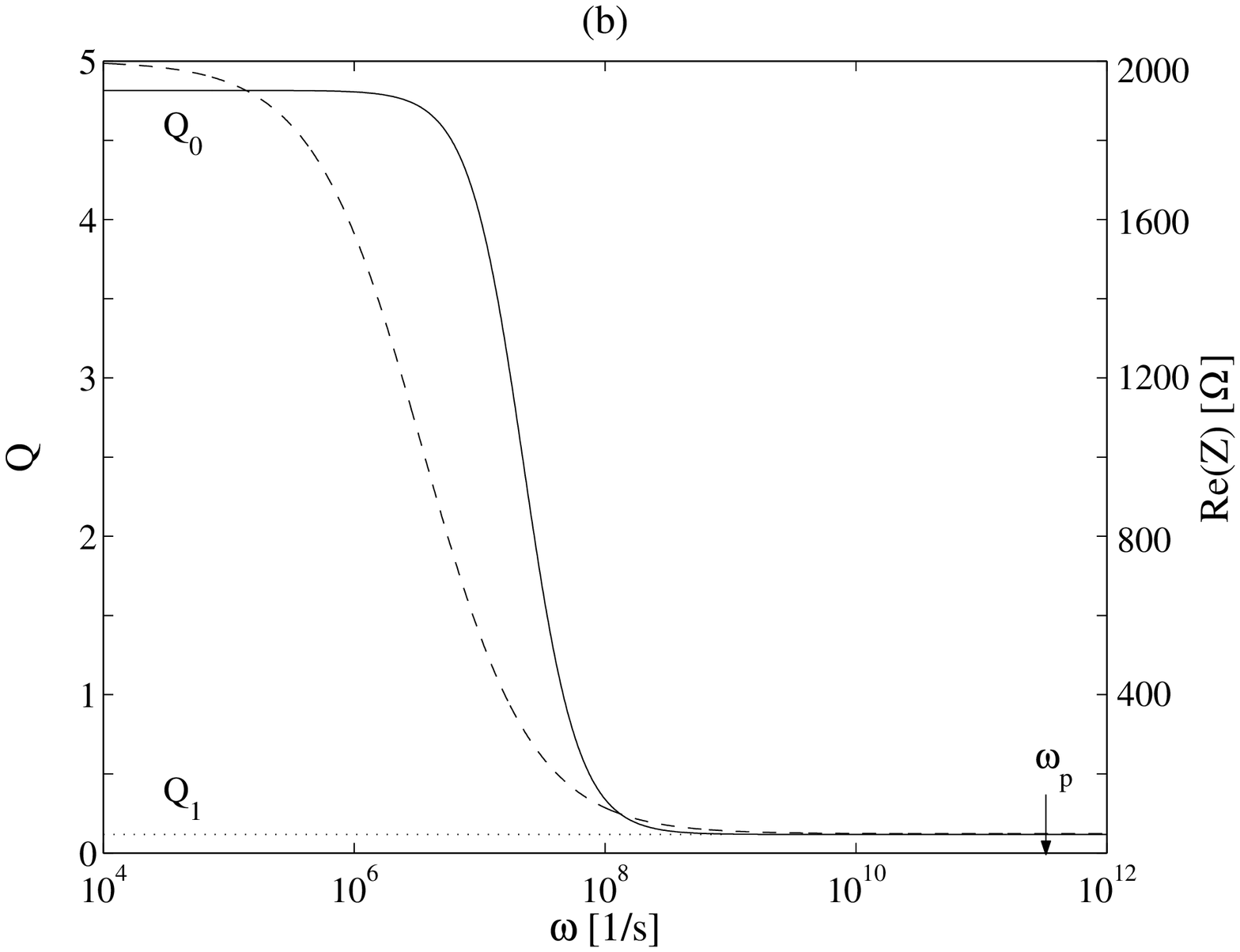}
\caption{(a) The bias pulse and hold, and (b) the frequency dependence of the shunt impedance $Re[Z]$ (dashed line) and of the quality factor $Q$ (solid line).}
\label{Curr_pulse}
\end{center}
\end{figure}
The bias pulse, which is shown in Fig.~\ref{Curr_pulse}(a), consists of a short switch pulse of amplitude $I_p$ and duration $\tau _p$, followed by a long hold level with amplitude $I_{hold}$ and duration $\tau _{hold}$. The idea behind this pulse is to quickly accelerate the phase particle above the critical velocity for dynamical switching, in order to force the switching process to happen as quickly as possible. The hold level is used to maintain the free running state with high voltage, long enough so that the voltage can be measured with an amplifier at room temperature. The hold level must be set high enough so that the phase particle will not be retrapped in the phase diffusion state, and yet low enough not to induce a late switching event. This type of bias pulse exploits the latching nature of the circuit to realize a sample-and-hold measurement strategy, in order to overcome measurement bandwidth limitations resulting from filters, lossy cables, and high gain amplifiers. The latch gives us the added advantage of having a binary detection of switching, with two distinct voltage outputs for switch or no-switch, from one bias pulse.  

The frequency dependent damping used in our numerical simulations is shown in Fig.~\ref{Curr_pulse}(b), where we have plotted the real impedance $Re[Z]$ of the bias circuit (dashed line), and the $Q$ factor of the junction (solid line) as function of frequency (see sec.~\ref{Exp} and \ref{Model} for junction and model parameters). The model used for the shunting impedence is a simple RC shunt (see sec.~\ref{Model}) which reproduces the qualitative feature of the actual experimental situation--that of overdamped phase dynamics at high frequency ($Q \ll 1$), crossing over to underdamped phase dynamics at low frequency ($Q \gg 1$). The frequency scale associated with the switch pulse ($1/ \tau_p$) can in principle be adjusted so that the junction is overdamped, and no hysteresis or latching behaviour of the circuit would result for the switch pulse alone. However, the frequency scale of the hold pulse ($1/ \tau_{hold}$) is such that underdamped dynamics is realized, and a hysteresis or latching of the junction voltage is possible.



\section {The Experiment}
\label{Exp}
\begin{figure}[tbp]
\begin{center}
\includegraphics[angle=0,width=0.42\textwidth]{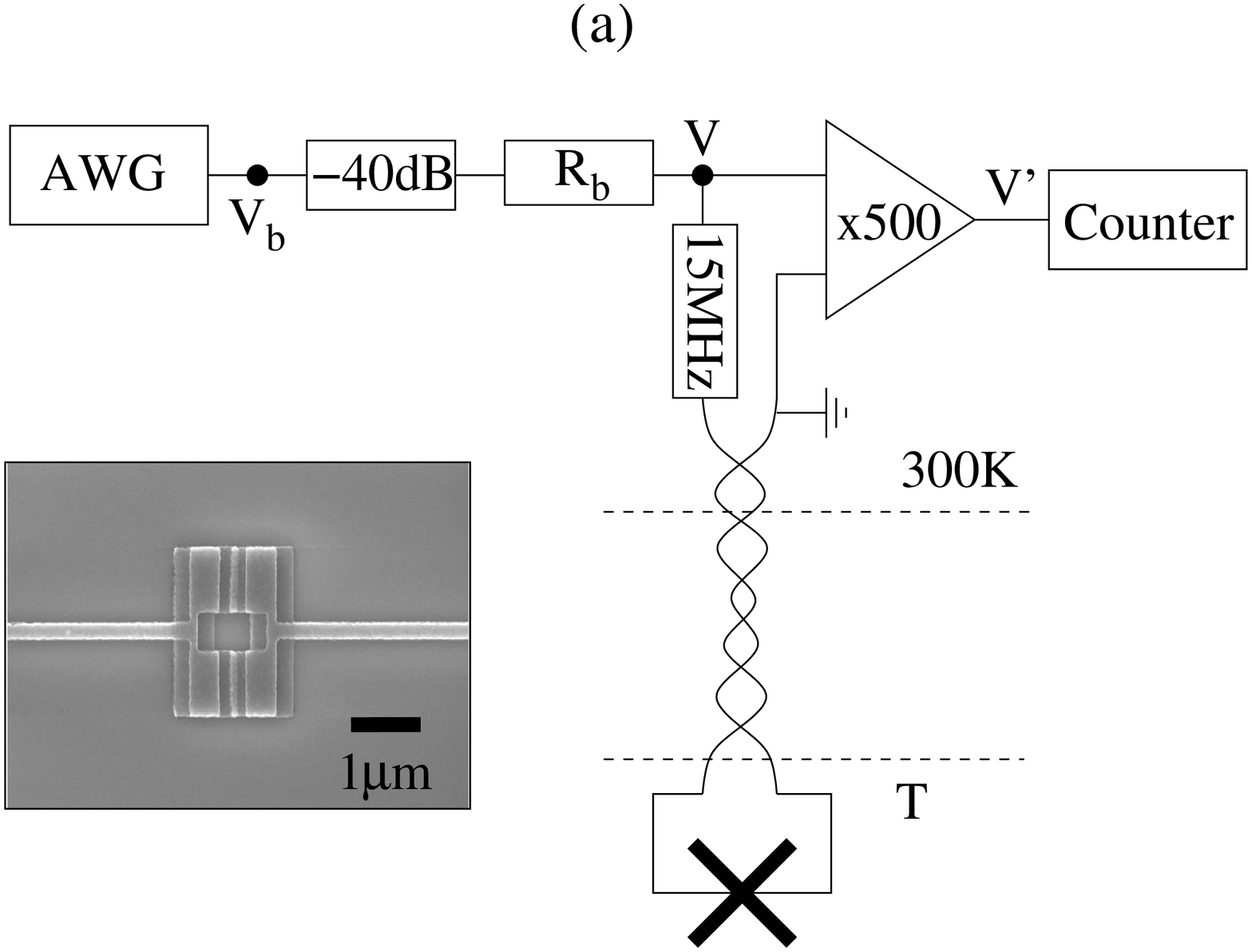}
\includegraphics[angle=0,width=0.33\textwidth]{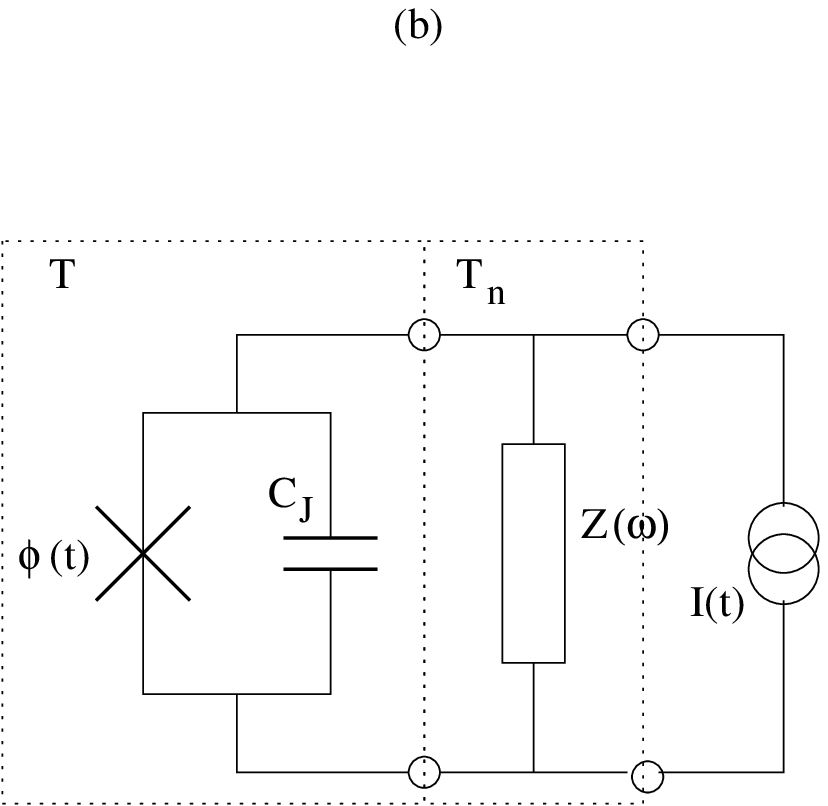}
\includegraphics[angle=0,width=0.19\textwidth]{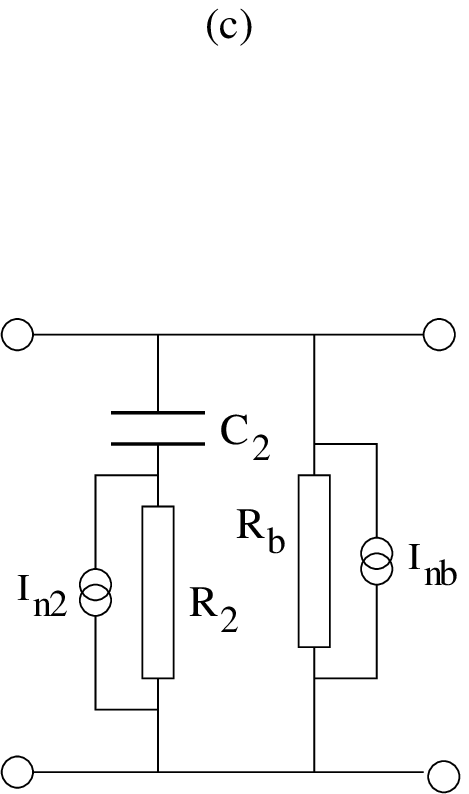}
\caption{(a) The experimental setup and sample shown in the inset. (b) The ideal version of the system consists of the Josephson junction (modelled by an ideal part and a capacitor $C_J$), connected to a current-source via some impedance $Z(\omega)$. (c) The impedance $Z$ we use. Note that we have included the noise currents due to the resistors.}
\label{Setup}
\end{center}
\end{figure}

\noindent
The sample used in the experiments is actually a superconducting quantum interference device (SQUID), which acts as an effective single junction and allows us to continuously tune the critical current, $I_0$, from its maximum value to zero by applying a magnetic field perpendicular to the loop. Here, however, we have only simulated data taken at a single value of $I_0$. The SQUID was fabricated by two angle evaporation of Al through a shadow mask which was defined by e-beam lithography. A micrograph of the sample is shown in the inset of Fig.~\ref{Setup}(a) together with a schematic of the experimental set up. An arbitrary waveform generator (AWG) biases the sample through an attenuator, a bias resistor via twisted pair cables of konstantan wire of length 2.1 m and DC resistance of 135 $\Omega$. The twisted pair acts as a lossy transmission line with a high frequency impedance of $\sim$ 50 $\Omega$. We have measured this value at frequencies up to 500 MHz. The losses in the line are such that reflections or standing waves were only very weakly visible. At higher frequency the actual impedance seen by the junction depends critically on the geometry of the sample mount between the leads and the junction, which we have not measured or modeled here. No additional cold filters were added at low temperature. The low frequency impedance seen by the junction is given by the bias resistor, $R_b=2$ k$\Omega$. The voltage across the sample is measured at the biasing end of the twisted pair cable through a gain 500 amplifier with bandwidth 100 kHz. The limited bandwidth of this amplifier does not allow us to observe the fast rise of the voltage when the junction switches to the free running state, the simulations however allow us to model the junction voltage in time. The bias resistor is chosen so that the voltage which builds up on the junction is kept well below the gap voltage $V_{2\Delta}=400 \ \mu$eV, in order to avoid quasiparticle tunneling in the junction and the associated dissipation. A digital counter registers a switching event when the output of the amplifier exceeds a trigger level. The switching probability is measured by counting the number of switching events and dividing by the number of applied bias pulses (typically $10^4$).\cite{JW}

The capacitance of the SQUID junctions (parallel combination) is $C_J \approx 7.2$fF, as estimated from the measured junction area and the specific capacitance of $45$ fF$/\mu m^2$. This gives a charging energy $E_C=4e^2/2 C_J = 45 \ \mu$eV, or $E_C/k_B=0.52$ K. The normal state resistance of the junction is 1.18 k$\Omega$ giving a critical current $I_0=265$ nA and Josephson energy $E_J = \hbar I_0/2e =540 \ \mu$eV, or $E_J/k_B=6.3$ K, or $E_J/E_C \approx 12$. This large ratio, together with the low tunneling impedance of the junction,
means that we are justified in treating the phase as a classical variable. Measurements were made in a dilution refrigerator with base temperature of 25 mK, well below the superconducting transition temperature of Al, $T_c \approx 1.2$ K. The temperature is varied in the range $0.025 \leq T \leq 0.7$ K as measured by a ruthenium oxide resistive thermometer in thermal equalibrium with the Cu sample mount. However, the effective temperature describing the noise in the theoretical model, is expected to be higher than the measured temperature as this noise is generated by dissipative elements located at higher temperatures.

The AWG is limited to a rise-time $\tau_{rise}=25 \ ns$. We have measured that this rise time is effectively transmitted to the sample, although some of the switch pulse amplitude is lost for the shortest pulses due to dispersion in the twisted pairs. The switch pulse has magnitude $I_{p}$ and length $\tau_p$ (including $2\tau_{rise}$) followed by a hold level of magnitude $I_{hold}<I_p$ and duration $\tau_{hold}$. Every such bias pulse is followed by a wait time $\tau_w$, with $I=0$, where the phase retraps and the junction returns to thermal equilibrium.  

\section{The model}
\label{Model}
\noindent
A model circuit for the measurement set up, shown in Fig.~\ref{Setup}(b), consists of an ideal Josephson junction with the current and voltage given by the Josephson relations, in parallel with the ideal capacitance of the tunnel junction $C_J$. This ideal tunnel junction is biased by a linear circuit with impedance $Z(\omega)$ in parallel with an ideal current source. (The model could have equivalently been cast in terms of a series combination of an impdeance and voltage source). The model used for the impedance consists of a series combination of a resistor $R_2=50 \Omega$ and capacitor $C_2 =0.14$ nF, in parallel with the bias resistor $R_b=2$ k$\Omega$ (shown in Fig.~\ref{Setup}(c)). With these values we match the measured impedance of the twisted pair fairly well. This type of model, with high impedance for low frequencies and low impedance for high frequencies (Fig.~\ref{Curr_pulse}(b)), was first proposed by Ono \textit{et al.}\cite{Ono87}, and it is the perhaps simplest possible model of a frequency dependent environment that is known to correctly model the qualitative features of the switching process. The sample is at a low temperature ($\Gamma=k_B T/E_J$ in dimensionless form) and the resistors are at an unknown noise temperature, $\Gamma_n=k_B T_n/E_J$, which give rise to the noise currents $I_{nb,2}$.

Applying Kirchoff's rules and the Josephson relations to the circuit in Fig~\ref{Setup}(b), one obtains a differential equation for $\phi(t)$. Defining the quantities $t'=t/t_s$ where $t_s=\hbar/2eI_0 Z_0$ and $Z_0=Z(0)$, $\omega'=t_s \omega$, $i=I/I_0$ (we will throughout this text use upper case $I$ for currents in amps and lower case $i$ for (dimensionless) currents in units of $I_0$) and $Q_0= Z_0 \sqrt{2eI_0 C_J / \hbar}$, one can write the equation in dimensionless form,
\begin{equation}
\label{Deq}
Q_0^2 \frac{d^2\phi}{dt'^2} = -\frac{du_i}{d\phi} + i_n - Z_0\int_0^{\infty} \frac{d\phi}{dt'}(t'-\tau') y(\tau') d\tau',
\end{equation}
where $u_i(\phi) = -\cos \phi - i \phi$ is the tilted washboard potential, $i_n$ is the noise current due to $Z$ with a correlation function obeying the fluctuation-dissipation theorem \cite{Ing} $\langle i_n(t'_1) i_n(t'_2) \rangle = \Gamma_n \frac{Z_0}{\pi} \int_{-\infty}^{+\infty} Re Y(\omega') \cos \omega' (t'_1-t'_2) d\omega'$, and the last term is the friction term in the form of a convolution integral between the voltage and the Fourier transform of the admittance, $y(t')=\frac{}{}\int_{-\infty}^{+\infty} Y(\omega') e^{i\omega' t'} d\omega'$ with $Y=1/Z$. Note that if $Z=R$, one recovers the RCSJ model.

Using our specific model for $Z$, we can write Eq.~(\ref{Deq}) as three coupled first order differential equations--a form more suitable for numerical analysis. We find
\begin{eqnarray}
\label{Deq_Zrrc}
\frac{d\phi}{dt'} & = & v \nonumber \\
\frac{dv}{dt'} & = & \frac{1}{Q_0^2} \Big[ i + i_{nb} + i_{n2} -\sin \phi -v + (v_C-v)(Q_0/Q_1 - 1) \Big] \\
\frac{d v_C}{dt'} & = & \frac{\rho}{Q_0^2} \Big[ v - v_C - \frac{i_{nb}}{(Q_0/Q_1-1)} \Big], \nonumber
\end{eqnarray}
where we have defined the following dimensionless parameters;
\begin{eqnarray}
& & v=\frac{V}{R_b I_0}, \ \ v_C=\frac{V_C}{R_b I_0}, \ \ \rho = \frac{R_b C_J}{R_2 C_2}, \nonumber \\
& &  \ \ Q_0=R_b\sqrt{\frac{2eI_0C_J}{\hbar}}, \ \ Q_1=\frac{1}{1/R_b+1/R_2}\sqrt{\frac{2eI_0C_J}{\hbar}}, \nonumber
\end{eqnarray}
where $v$ is the dimensionless voltage over the junction and $v_C$ is the dimensionless voltage over $C_2$. $i_{nj}$ is the noise current due to resistor $R_j$ ($j=b,2$), with correlation function $\langle i_{nj}(t'_1) i_{nj}(t'_2) \rangle = 2\Gamma_n \frac{R_b}{R_j} \delta(t'_2 - t'_1)$ where we for simplicity have assumed that $R_b$ and $R_2$ are at the same temperature. The generic features of the solutions to Eq.~(\ref{Deq_Zrrc}) are known and have been investigated thoroughly by e.g. Kautz and Martinis.\cite{KM90}

The frequency dependent quality factor, plotted in Fig.~\ref{Curr_pulse}(b), is
\begin{equation}
\label{Qow}
Q(\omega)=Q_0 \frac{1+Q_0^2 \rho^{-2}(\omega/\omega_p)^2}{1 + Q_0^3 Q_1^{-1}\rho^{-2}(\omega/\omega_p)^2}
\end{equation}
where $\omega_p=\sqrt{2eI_0/\hbar C_J}$ is the plasma frequency (indicated in Fig.~\ref{Curr_pulse}(b)). This expression is derived using the following definition; $Q(\omega)=\sqrt{2eI_0C_J/\hbar G^2(\omega)}$ where $G=Re[1/Z]$. 

We solve Eq.~(\ref{Deq_Zrrc}) using a 4'th order Runge-Kutta algorithm. The noise curents are $i_{nb,2}=N \sigma_{b,2}$, where $N \in [-1,1]$ is a Gaussian distributed random number, $\sigma_{b,2}= \sqrt{\frac{2\Gamma_n}{\Delta t'}\frac{R_b}{R_{b,2}}}$ is the standard deviation and $\Delta t'$ is the dimensionless timestep in the numerical routine. This last factor is present since it is the mean value of $i_{nb,2}$ during a time $\Delta t'$ that enters Eq.~(\ref{Deq_Zrrc}).

We have simulated the switching process using $\tau_p=0.1, 1.0$ and $10 \ \mu s$. If the junction switches, there will be a finite voltage $\langle V \rangle > V_{trig}$ over the junction (where $\langle V \rangle$ denotes the mean value over time and $V_{trig}$ is the trigger voltage) and the counter will click. The scale of $\langle V \rangle$ is determined by $R_b$ and $I_{hold}$. The finite voltage is latched for the duration of the hold pulse due to the hysteretic behaviour of the non-linear circuit. In the no-switch case, the phase particle either fluctuates thermally ($\langle V \rangle=0$) or diffuses down the washboard ($\langle V \rangle< V_{trig}$)—-in both cases the counter will not click, assuming $V_{trig}$ is chosen high enough. 


\section{Results and discussion}
\begin{figure}[tbp]
\begin{center}
\includegraphics[angle=0,width=0.48\textwidth]{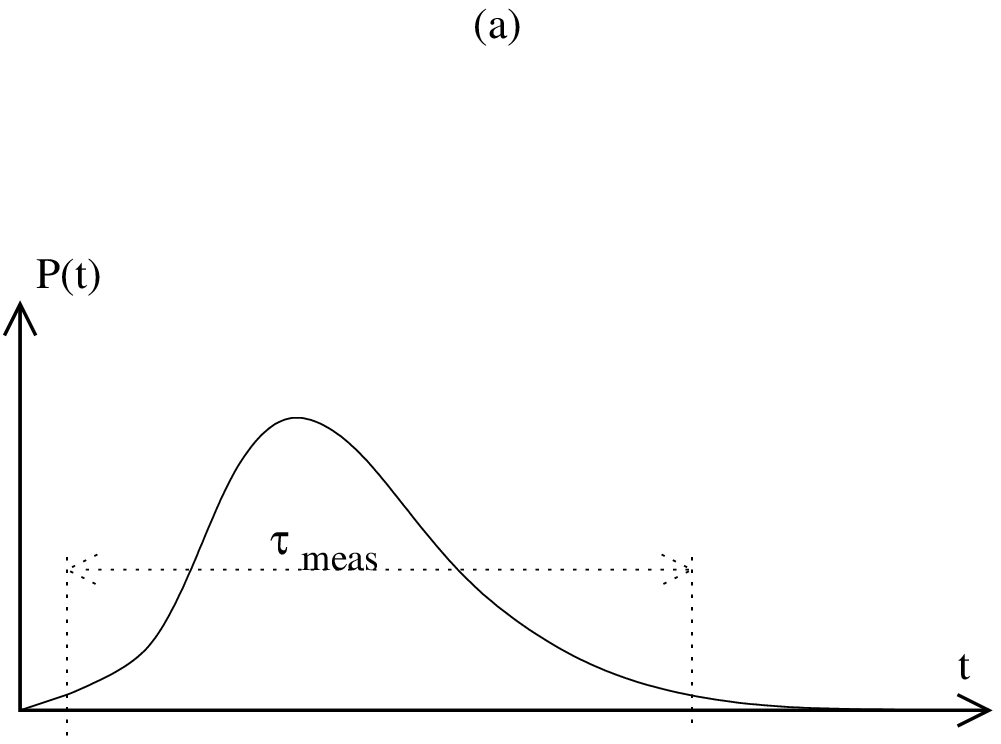}
\includegraphics[angle=0,width=0.40\textwidth]{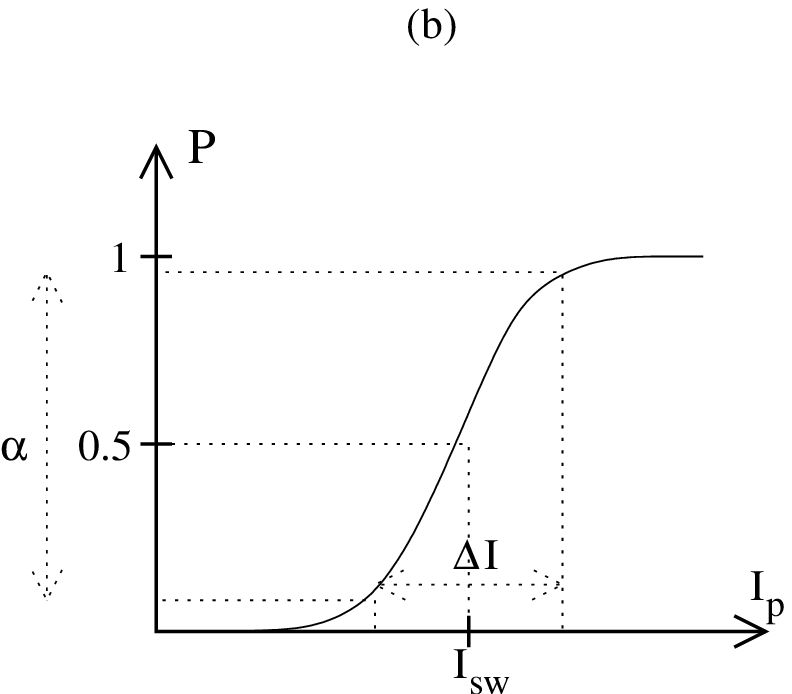}
\caption{(a) A probability distribution of switching as a function of time, with the measurement time defined as the width of $98\%$ of this distribution. (b) A graphical representation of the discriminating power $\alpha$ and the current resolution $\Delta I$.}
\label{Res_pic}
\end{center}
\end{figure}

\noindent
Two important quantities for the use of this switching process as a detector of the quantum state of a circuit, are the measurement time and the resolution. The measurement time can be determined by recording when the switching events occur, i.e. at what time the simulated voltage reaches the prescribed trigger level. Due to bandwidth limitations in the actual experiment, we can not measure this time. However, we can simulate many switching events to determine the probability, $P(t)dt$, that the switching event occurs between the time $t$ and $t+dt$. A typical distribution $P(t)$ is shown in Fig.~\ref{Res_pic}(a). We define the measurement time to be the width of 98\% of this distribution, neglecting the first and last percent. The resolution of the detector is determined from the probability, $P(I_p)dI_p$, that a switching event will occur with a switch pulse of amplitude $I_p$. This probability distribution can be both simulated and measured, and a typical curve is shown in Fig.~\ref{Res_pic}(b). The resolution of our detector $\Delta I$, is the ability of the detector to discriminate between two different values of the switching current (which will differ depending on the quantum state of a circuit). We have (arbitrarily) chosen a discriminating power $\alpha=0.8$ (meaning that 20\% of the events will be miscounted) to determine the resolution $\Delta I$.

\begin{figure}[tbp]
\begin{center}
\includegraphics[angle=0,width=0.55\textwidth]{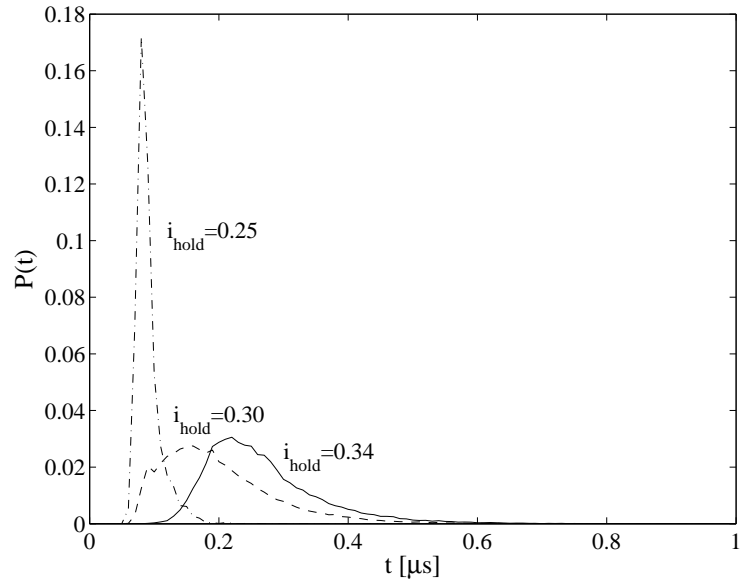}
\caption{This shows $P(t)$ during one pulse, for $dt=0.01 \ \mu$s, $P=0.50$ and $V_{trig}=35 \ \mu$V, with $\Gamma_n=0.47$ and $i_{hold}=0.34$ (solid line), $i_{hold}=0.30$ (dashed line) and $i_{hold}=0.25$ (dash-dotted line).}
\label{Sw_ProbDist}
\end{center}
\end{figure}

Fig.~\ref{Sw_ProbDist} shows a simulated $P(t)$ for three different values of the hold level, $i_{hold}$. The magnitude of the noise was set to $\Gamma_n=k_B T_n/E_J =0.47$ (corresponding to $T_n=3.0$ K with $E_J$ from sec.~\ref{Exp}) and the voltage trigger to $V_{trig}=35 \ \mu$V, which is safely above the phase diffusion voltage. The switch pulse had duration $\tau _p = 0.1 \ \mu$s and the hold pulse had duration $\tau_{hold}=0.9 \ \mu$s. The amplitude of the switch pulse $i_p$ was adjusted for each curve so that $P(I_p)=0.5$. In this simulation it was first determined that a pulse consisting of the hold level itself (i.e. when $i_p=i_{hold}$) did not induce any switching ($P<0.001$). We can see from the solid curve of Fig.~\ref{Sw_ProbDist} that the hold level $i_{hold}=0.34$ induces late switching events, with the actual switching taking place during the hold time, resulting in a measurement time that exceeds $\tau _p$. Lowering the hold level to $i_{hold}=0.3$  (dashed curve of Fig. \ref{Sw_ProbDist}) causes the switches to occur earlier in time, with a slight reduction of $\tau_{meas}$. If the hold level is lowered to $i_{hold}=0.25$ (dash-dotted curve of Fig.~\ref{Sw_ProbDist}) we find that the distribution $P(t)$ becomes sharply peaked, with $\tau_{meas} \approx \tau_p$.  

The simulations clearly show that $\tau_{meas}$ increases with $i_{hold}$. We find $\tau_{meas}=0.10, 0.42, 0.44 \ \mu$s for $i_{hold}=0.25, 0.30, 0.34$ respectively. These values of $\tau_{meas}$ are approximately the same for different switching probabilities in the region $i_p=i_{sw} \pm \Delta i/2$. For $i_{hold}>0.34$, $\tau_{meas}$ will increase, since the hold pulse alone will begin to induce switching events. For $i_{hold}<0.25$, $\tau_{meas}$ will continue to be close to $\tau_p$, although for very low $i_{hold}$ it will become hard to distinguish between switch and no-switching events. We have also run this simulation with lower strength of the noise term, $\Gamma_n=0.11$, and we observe $\tau_{meas} \approx 0.1-0.15 \ \mu$s for $0.40 \leq i_{hold} \leq 0.63$. As expected, a weaker noise term will give a more rapid measurement time over a wider range of hold levels.  

From this simulation we can conclude the following: 1) Even in the presence of significant noise ($\Gamma _n=0.47$) it is possible to find a switch pulse amplitude and hold level which give a rapid dynamical switching that occurs during the switch pulse without a significant number of late switches. 2) Simply adjusting the hold level for zero switching when $i_p=i_{hold}$ is not good enough to achieve optimal measurement time. Because one can not observe the escape process in the experiment, simulations of this kind are necessary to verify that the hold level is sufficiently low.    

\begin{figure}[tbp]
\begin{center}
\includegraphics[angle=0,width=0.57\textwidth]{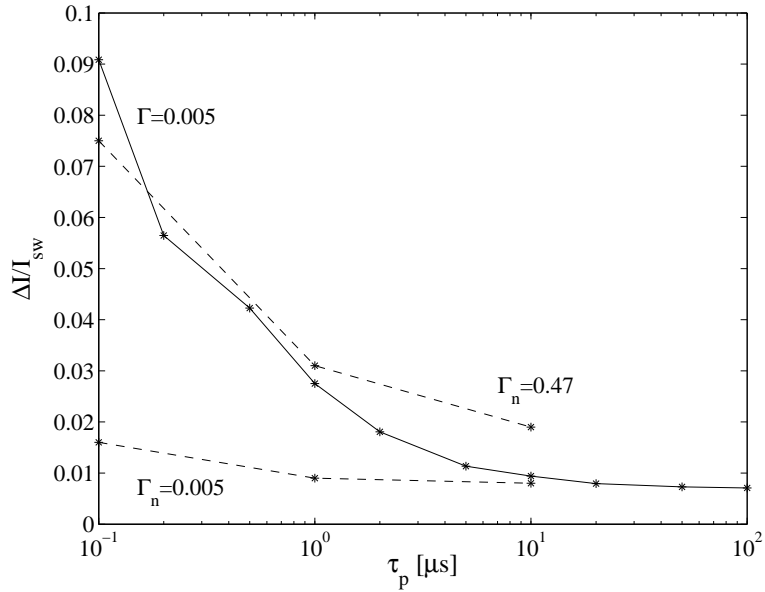}
\caption{This shows $\Delta I/I_{sw}$ as a function of $\tau_p$. The dashed curves are simulated data with $\Gamma_n=0.005$ and $0.47$, and the solid is measured data with $\Gamma=0.005$ ($T=0.030$ K).}
\label{Di_data}
\end{center}
\end{figure}

Several switching curves $P(I)$ of the type shown schematically in Fig.~\ref{Res_pic}(b) were simulated for various values of the temperature and switching pulse parameters, as they were adjusted in the actual experiment. In Fig.~\ref{Di_data} we show a comparison between the simulated and measured resolution $\Delta I/I_{sw}$, as a function of the switch pulse duration $\tau_p$. Here, $I_{sw}$ is the value of the switch pulse amplitude which gives 50\% switching probability, $P(I_p)=0.5$. It is difficult to compare the simulated and experimental value of $I_{sw}$, because in the experiment we measure the voltage pulse applied to the bias resistor. Dispersion in the twisted pairs, and parasitic capacitance of the sample mount, make it difficult to know the actual amplitude of the current pulse applied to the junction.  

The simulated values shown in Fig.~\ref{Di_data} are calculated for two different noise tempertures. The value $\Gamma_n=0.005$ (or $T_n=30$ mK for the particular junction parameters considered), corresponds to the measured temperature of the sample mount during the experiment. For this value of the noise temperature we see that the experimental and simulated values agree for longer $\tau_p$, while for shorter pulses the simulation gives much higher resolution than the experiment. We have adjusted the noise temperature, and we find reasonable agreement over the range of $\tau _p$ studied for  $\Gamma_n=0.47$ ($T=3$ K). This temperature seems excessively high, but note that in the simulation we have put $R_b$ and $R_2$ at the same temperature. It would be more realistic to take $R_b$ at a higher, and $R_2$ at a lower temperature. In turn, this would increase $i_{nb}$ and decrease $i_{n2}$ in Eq.~(\ref{Deq_Zrrc}), causing a partial cancellation effect, giving an intermediate effective temperature. The exact consequences remains though to be investigated.

Both the experimental and theoretical curves of Fig.~\ref{Di_data} show decreasing resolution (increasing $\Delta I$) for shorter measurement time (smaller $\tau_p$). Partly this is due to that $i_{sw}$ increases with decreasing $\tau_p$ for given $\Gamma_n$, but also the statistical fluctuation of the mean noise amplitude should be bigger for a short pulse than for a long pulse. Thus, as expected, there is a trade off between measurement speed and measurement resolution. 

\section{Summary}
\noindent
In summary, we have performed computer simulations of the dynamical switching process of a Josephson junction subject to noise and to frequency dependent damping. The model for damping is appropriate for measurements of small capacitance Josephson junctions, where the phase dynamics is overdamped at high frequencies, and underdamped at low frequencies. We use a special bias current pulse, consisting of a short switch pulse followed by a longer hold level, appropriate for binary detection of the switching process. We characterize the switching process in terms of speed, $\tau_{meas}$, and resolution, $\Delta I/I_{sw}$. The simulations show that it is possible to achieve a minimum switching time given by the duration of the fast switching pulse, even in the presence of strong noise. To achieve this fast measurement, the hold level must be set appropriately low. As expected, there is a trade off between measurement speed and resolution.

\begin{center}
\Large{\bf Acknowledgments }\\
\end{center}
We wish to thank D. Vion for interesting discussions. This work was partially supported by the EU project SQUBIT, in the 5th framework QIPC program, by the Swedish Research Council and by the Swedish SSF. \\

\end{document}